\documentclass[a4paper,12pt]{article}
\usepackage{epsfig}
\usepackage{amssymb}
\usepackage{amsfonts}
\usepackage{amsmath}
\usepackage{euscript}
\usepackage{verbatim}
\usepackage{latexsym}
\usepackage{graphicx}

	\usepackage{graphicx}	
	\usepackage{subfigure}
	\usepackage{wrapfig}
	\usepackage[T1]{fontenc}
	\usepackage{mathtools}
	\usepackage{float}

	\usepackage{braket}
	\usepackage{caption}

\usepackage{tikz}
\usetikzlibrary{shapes.geometric, arrows,patterns,snakes}
\tikzstyle{ellip} = [ellipse, minimum width=3cm, minimum height=1cm,text centered, draw=black]

\newskip\humongous \humongous=0pt plus 1000pt minus 1000pt

\newif\ifdtup

\catcode`\@=11

\@addtoreset{equation}{section}

\def\@normalsize{\@setsize\normalsize{15pt}\xiipt\@xiipt
\abovedisplayskip 14pt plus3pt minus3pt%
\belowdisplayskip \abovedisplayskip
\abovedisplayshortskip \z@ plus3pt%
\belowdisplayshortskip 7pt plus3.5pt minus0pt}

\def\small{\@setsize\small{13.6pt}\xipt\@xipt
\abovedisplayskip 13pt plus3pt minus3pt%
\belowdisplayskip \abovedisplayskip
\abovedisplayshortskip \z@ plus3pt%
\belowdisplayshortskip 7pt plus3.5pt minus0pt
\def\@listi{\parsep 4.5pt plus 2pt minus 1pt
     \itemsep \parsep
     \topsep 9pt plus 3pt minus 3pt}}

\relax

\catcode`@=12

\catcode`\@=11

\def\section{\@startsection{section}{1}{\z@}{3.5ex plus 1ex minus
   .2ex}{2.3ex plus .2ex}{\large\bf}}

\def\SymBoxes#1#2#3#4{\newdimen\un@t \un@t#3%
\raisebox{#1}{\rule{#2\un@t}{#4}\hskip-#2\un@t
\@tempdimb\un@t \advance\@tempdimb by-#4\@tempcntb#2\relax%
\@whilenum{\@tempcntb>0}\do{
\rule{#4}{\un@t}\hskip\@tempdimb \advance\@tempcntb by\m@ne}%
\hskip-#2\un@t \rule[\un@t]{#2\un@t}{#4}%
\rule[\un@t]{#4}{#4}\hskip-#4
\rule{#4}{\un@t}}\hskip-#4}                

\begin{document}

\newcommand{\beq}{\begin{equation}}
\newcommand{\eeq}{\end{equation}}
\newcommand{\bea}{\begin{eqnarray}}
\newcommand{\eea}{\end{eqnarray}}
\newcommand{\beas}{\begin{eqnarray*}}
\newcommand{\eeas}{\end{eqnarray*}}
\newcommand{\defi}{\stackrel{\rm def}{=}}
\newcommand{\non}{\nonumber}
\newcommand{\bquo}{\begin{quote}}
\newcommand{\enqu}{\end{quote}}
\renewcommand{\(}{\begin{equation}}
\renewcommand{\)}{\end{equation}}
\def \eqn#1#2{\begin{equation}#2\label{#1}\end{equation}}

\def\Fr{\left(1-\frac{2M}{r}\right)}
\def\Frv{\left(1-\frac{2M(v)}{r}\right)}
\def\e{\epsilon}
\def\IZ{{\mathbb Z}}
\def\IR{{\mathbb R}}
\def\IC{{\mathbb C}}
\def\IQ{{\mathbb Q}}
\def\de{\partial}
\def\Tr{ \hbox{\rm Tr}}
\def\H{ \hbox{\rm H}}
\def\HE{ \hbox{$\rm H^{even}$}}
\def\HO{ \hbox{$\rm H^{odd}$}}
\def\K{ \hbox{\rm K}}
\def\Im{ \hbox{\rm Im}}
\def\Ker{ \hbox{\rm Ker}}
\def\const{\hbox {\rm const.}}
\def\o{\over}
\def\im{\hbox{\rm Im}}
\def\re{\hbox{\rm Re}}
\def\bra{\langle}\def\ket{\rangle}
\def\Arg{\hbox {\rm Arg}}
\def\Re{\hbox {\rm Re}}
\def\Im{\hbox {\rm Im}}
\def\exo{\hbox {\rm exp}}
\def\diag{\hbox{\rm diag}}
\def\longvert{{\rule[-2mm]{0.1mm}{7mm}}\,}
\def\a{\alpha}
\def\dag{{}^{\dagger}}
\def\tq{{\widetilde q}}
\def\p{{}^{\prime}}
\def\W{W}
\def\N{{\cal N}}
\def\calB{\mathcal{B}}
\def\hsp{,\hspace{.7cm}}

\def\br{\nonumber\\}
\def\IZ{{\mathbb Z}}
\def\IR{{\mathbb R}}
\def\IC{{\mathbb C}}
\def\IQ{{\mathbb Q}}
\def\IP{{\mathbb P}}
\def \eqn#1#2{\begin{equation}#2\label{#1}\end{equation}}

\newcommand{\M}{\ensuremath{\mathcal{M}}
                    }
\newcommand{\oc}{\ensuremath{\overline{c}}}
\begin{titlepage}
\begin{flushright}
CHEP XXXXX
\end{flushright}
\bigskip
\def\thefootnote{\fnsymbol{footnote}}

\begin{center}
{\Large
{\bf A Dynamical Boundary for Anti-de Sitter Space \\ 
 \vspace{0.1in} 
}
}
\end{center}

\bigskip
\begin{center}
{
Chethan KRISHNAN$^a$\footnote{\texttt{chethan.krishnan@gmail.com}}, Avinash RAJU$^a$\footnote{\texttt{avinashraju777@gmail.com}}, P.N. Bala SUBRAMANIAN$^a$\footnote{\texttt{pnbalasubramanian@gmail.com}} }
\vspace{0.1in}

\end{center}

\renewcommand{\thefootnote}{\arabic{footnote}}

\begin{center}

$^a$ {Center for High Energy Physics,\\
Indian Institute of Science, Bangalore 560012, India}\\

\end{center}

\noindent
\begin{center} {\bf Abstract} \end{center}
We argue that a natural boundary condition for gravity in asymptotically AdS spaces is to hold the {\em renormalized} boundary stress tensor density fixed, instead of the boundary metric. This leads to a well-defined variational problem, as well as new counter-terms and a finite on-shell action. We elaborate this in various (even and odd) dimensions in the language of holographic renormalization. Even though the {\em form} of the new renormalized action is distinct from the standard one, once the cut-off is taken to infinity, their {\em values} on classical solutions coincide when the trace anomaly vanishes. For AdS$_4$, we compute the ADM form of this renormalized action and show in detail how the correct thermodynamics of Kerr-AdS black holes emerge. We comment on the possibility of a consistent quantization with our boundary conditions when the boundary is dynamical, and make a connection to the results of Compere and Marolf. The difference between our approach and microcanonical-like ensembles in standard AdS/CFT is emphasized.


\vspace{1.6 cm}
\vfill

\end{titlepage}

\setcounter{footnote}{0}


\section{Introduction} 

Historically, most of the work on boundary conditions in gravity has been in the context of Dirichlet boundary conditions: the Gibbons-Hawking-York (GHY) boundary term \cite{York} was the first boundary term to be identified that made the variational problem for gravity well-defined. It also gave a formal yet compelling basis for horizon thermodynamics \cite{GHY, CK}. In the usual AdS/CFT correspondence \cite{MaldacenaAdSCFT, GKP, Witten}, the boundary values of fields on the gravity side are identified as the sources of the fields in the field theory. Thus AdS/CFT correspondence is formulated as a Dirichlet problem as well (on the gravity side). 

Recently however a boundary term for gravity  \cite{KR} has been introduced (see also various related work \cite{Grumiller, padmanabhan, Vassilevich, Park, McNees, HairyBH,Geoffrey,Apolo,Olea,Papadimitriou,marc,cases}) which is a natural candidate for a Neumann formulation of gravity. Furthermore it was shown \cite{latest} that various thermodynamical aspects of gravity can in fact be reproduced using this Neumann boundary term as well. In light of this, in this paper, we will explore gravity in asymptotically Anti-de Sitter spacetimes with Neumann boundary conditions.

The proposal of \cite{KR} was to treat Neumann boundary condition as holding the canonical conjugate of the boundary metric\footnote{This turns out to be the boundary stress tensor density.} fixed. In particle mechanics and field theory, holding the canonical conjugate of the boundary value of the field fixed is identical to the usual Neumann boundary conditions, but in gravity this leads to an alternative to holding the normal derivative\footnote{To the best of our knowledge, a boundary term with the normal derivative at the boundary fixed, is not known for gravity.} of the boundary metric fixed, and leads to a well-defined new boundary term \cite{KR}. The translation from Dirichlet to Neumann can be understood as a Legendre transform \cite{latest, KW}. 

Typically, to get a finite action on solutions, one has to take care of infrared divergences of the Einstein-Hilbert action in both flat space and in AdS. This is true even with the addition of boundary terms that make the variational problem well-defined. In flat space, this was done for the GHY boundary term in \cite{GHY} and for the Neumann term in \cite{latest} via appropriate background subtraction procedures. In AdS however, for the Dirichlet problem, there exists a well-defined and quite natural way to get finite actions by the addition of counter-terms \cite{balasubramanian, deHaro}, which have a very natural interpretation in the dual field theory as canceling UV divergences. Such counter-terms lead to a finite action and a finite (renormalized) stress tensor.  

The existence of this finite stress tensor suggests that in AdS, one can define the Neumann variational problem to be one where we hold the renormalized stress tensor density fixed, and one should get a well-defined variational principle and finite Neumann action. We can do this in two ways: we can do this via starting from the renormalized Dirichlet action in AdS (which is well-known from, say, \cite{deHaro}) and do a Legendre transform on the boundary metric, or we can start from a Fefferman-Graham expansion as the definition of asymptotically AdS space, and systematically construct counter-terms for the un-renormalized Neumann action by demanding vanishing of divergences. In the next section, we will adopt the latter strategy and write down explicit renormalized Neumann actions in AdS$_{d+1}$ with $d=2,3,4$. Remarkably, we will find in an Appendix that both these approaches yield the same results. 

In a later section we will evaluate the finite actions that this leads to on classical (black hole) solutions. We will also find that these actions have the same numerical values as the corresponding Dirichlet actions on these solutions, upto subleading terms that vanish when the radial cut-off is taken to infinity. This is not surprising, because the Legendre transform relating the two actions is of the form
\bea
S^{ren}_N=S^{ren}_D-\int_{\partial {\cal M}}\ \pi^{ij}\gamma_{ij}
\eea
where $\gamma_{ij}$ is the boundary metric and $\pi^{ij}$ is its canonical conjugate, and is equal to the renormalized energy momentum tensor density. This means that $\pi^{ij}\gamma_{ij}$ is proportional to the trace of the boundary stress tensor and so when the conformal anomaly of the boundary theory vanishes, this object is zero on classical solutions\footnote{As it happens, since the conformal anomaly is related to the curvatures of the boundary surface, when these curvatures are vanishing, we will see a match for standard black hole solutions between Dirichlet and Neumann also in AdS$_{d+1}$ with even $d$.}.

We will see however that even though the values can be same (as cut-off is taken to infinity), the forms of the renormalized Dirichlet and Neumann actions can be quite different. To illustrate this in some detail, we will compute the ADM version of both these actions. We will also find that comparing the covariant and Hamiltonian ways of evaluating these actions yields the generalized Smarr formula, but in different ways. The covariant-canonical relations and the Smarr formula automatically imply the first law as well \cite{Klemm}.

Our results are conservatively thought of merely as a new boundary condition for classical AdS gravity with suitable boundary terms, but we find it plausible that our results go beyond classical. We believe they are indicative of a possibly interesting boundary condition for quantum gravity in AdS. This might seem a-priori impossible because consistent quantization of fields in AdS requires that they be normalizable (or finite energy), and except in some windows of masses \cite{KW} for scalars (say), it is known that only one boundary condition leads to consistent quantization for fields in a fixed AdS background. We will argue however that this is not quite true: the reason is that the notion of energy in a fixed AdS background is different from that in an AdS where the metric is dynamical \cite{MincesRivelles}. In particular we will speculate (partly inspired by a result of Compere and Marolf) that it may be possible that our boundary conditions are consistent at the quantum level when the boundary metric in AdS is dynamical: that is, the boundary theory contains dynamical gravity. We leave a conclusive take on this problem for later work.

Holding the boundary metric fixed is the standard way of thinking about AdS familiar from AdS/CFT. To clarify some points which might cause confusion, we conclude by elaborating a little on the choice of ensembles in AdS/CFT. We relegate some of the relevant facts we need (among other things) to the appendices.

\section{Holographic Renormalization of Neumann Gravity}

In this section we will derive the renormalized Neumann action by directly dealing with the Fefferman-Graham expansion (\ref{FG}) and demanding that the action be finite. Typically in Dirichlet theory one imagines that the boundary conditions are set by the leading part of the FG expansion, in our case it is a combination of the $g_i$'s (see (\ref{FG}) that is getting fixed via the renormalized boundary stress tensor. A standard review is \cite{skn}.

\subsection{Regularized Action in Fefferman-Graham Coordinates}

By asymptotically AdS$_{d+1}$ space, in this paper we will mean a metric that solves the Einstein equation with a negative cosmological constant, that can be expressed asymptotically (\textit{i.e.}, as $z \rightarrow 0$) by a general Fefferman-Graham expansion given by
\begin{equation}
ds^2 = G_{\mu \nu}dx^{\mu} dx^{\nu} = \frac{l^2}{z^2}\left( dz^2 + g_{ij}(x,z)dx^i dx^j \right)
\end{equation}
where
\begin{equation}
g(x,z) = g_{0} + z^2 g_2 + \cdots + z^d g_d + z^d \log z^2\; h_d + O(z^{d+1}).\label{FG}
\end{equation}
Only even powers of $z$ appear up to $O(z^{[d-1]})$. The log term appears only for even $d$. In all the discussions that follow, we set $l=1$. The cosmological constant is related to the AdS radius through the relation $\Lambda = - \frac{d(d-1)}{2l^2}$. Since only even powers appear in the expansion, we introduce a new coordinate $\rho = z^2$ in which the metric takes the form
\begin{eqnarray}\label{asymptotic_metric}
ds^2 &=& \frac{d\rho^2}{4\rho^2} + \frac{1}{\rho}g_{ij}(x,\rho)dx^i dx^j \\ \nonumber
g(x,\rho) &=& g_0 + \rho g_2 + \cdots + \rho^{d/2} g_d + \rho^{d/2}\;\log \rho \; h_d 
\end{eqnarray}
Note that the condition that this metric solves the Einstein equation means that the higher order $g_{(m)ij}$ can be determined in terms of the lower order ones, and explicit formulas can be written down for them. We present explicit expressions in an Appendix. 

We can compute\footnote{In what follows, $\gamma_{ij}$ is the induced metric on $\partial {\cal M}$ and  $\varepsilon$ takes values $\pm 1$ depending on whether  $\partial {\cal M}$ is time-like or space-like respectively. $\Theta $ is the trace of extrinsic curvature of $\partial {\cal M}$ which is defined to be $\Theta_{ij}=\nabla_{(a}n_{b)}e^a_i e^b_j$,  where $n_a$ is the outward pointing normal vector and $e^a_i = \frac{\partial x^a}{\partial y^i}$ is the projector arising from the bulk coordinates $x^a$ and the boundary coordinates $y^i$. } the Neumann action \cite{KR, latest} (note that \cite{KR} worked with the bulk dimension, so our $d=D-1$ in the notation there),
\begin{equation}\label{neumann_action}
S_{N} = \frac{1}{2\kappa}\int_{\M}d^{d+1}x \sqrt{-g}(R-2\Lambda) - \frac{(d-3)}{2\kappa}\int_{\partial \M}d^{d}y \sqrt{|\gamma|}\varepsilon \Theta
\end{equation}
for (\ref{asymptotic_metric}) and we immediately sees that it diverges. This is not a surprise: the same thing happens for the Dirichlet action as well, and the process of adding counter-terms to the Dirichlet action to make it finite is known as holographic renormalization \cite{deHaro}. We can adopt a similar approach here. The first step is to cut-off the radial integration at a finite $\rho=\epsilon$, to regulate the action. After this regularization, the Neumann action (\ref{neumann_action}) is given by
\begin{equation}
S_{N}^{reg} = -\frac{d}{2\kappa}\int d^d x \int_{\epsilon}d\rho \; \frac{1}{\rho^{d/2 +1}}\sqrt{-g} -\frac{(d-3)}{2\kappa}\int d^d x \left. \frac{1}{\rho^{d/2}}\left( d\sqrt{-g} - 2 \rho \partial_{\rho}\sqrt{-g} \right) \right|_{\rho = \epsilon}
\end{equation}

Our goal is to add counter-terms so that the Neumann action becomes finite. We will find that this is indeed a natural construction and for standard black hole solutions it leads to the same on-shell action as the Dirichlet theory.

\subsection{AdS$_3$ ($d=2$)}

In $d=2$ the regularized Neumann action takes the form,

\begin{eqnarray}
S_{N}^{reg} &=& -\frac{1}{\kappa}\int d^2 x \left[ \int_{\epsilon} d\rho \frac{\sqrt{-g}}{\rho^2} + \left. \left( -\frac{\sqrt{-g}}{\epsilon} + \partial_{\rho}\sqrt{-g} \right) \right|_{\rho = \epsilon} \right]
\end{eqnarray}
Using the expansion for the determinant \eqref{detexpansion} and doing the $\rho$ integral, we arrive at following final form for the regulated action

\begin{equation}
S^{reg}_N = \frac{1}{2\kappa}\int d^2 x \sqrt{-g_0} \log \epsilon \; \Tr\; g_{2} \label{2log}
\end{equation}
In this paper, we will ignore this Logarithmic divergence, because it will not be relevant for the situations we consider, like black holes. This is similar to the approach of \cite{balasubramanian} and we would like to write down counter-terms parallel to theirs in terms of the induced metric. The logarithmic divergence in the Dirichlet case were presented later in \cite{deHaro}. We emphasize however that even though we do not use them, our presentation of logarithmic divergences is complete: the expressions for the quantities involving $g_2$ in \eqref{2log}, \eqref{4log} in terms of curvatures of the boundary metric $g_0$ are presented in an Appendix. Note however that unlike the other counter-terms, we cannot absorb the cut-off dependence of the logarithmic divergence entirely into expressions involving the induced metric; a logarithmic cut-off dependence will remain. This is unavoidable, and this is the form in which \cite{deHaro} also leave their results, see their equation (B.4), last term. The renormalized quantities are of course cut-off independent by construction. 

Once we ignore the logarithmic term, the renormalized Neumann action is therefore identical to the original Neumann action $S_{N}$ in three dimensions: no counter-terms are required to render the action finite.  
\bea
S^{ren}_N=S_N
\eea
This was an observation that was already made in a slightly different language in \cite{Grumiller, BanadosMendez}, as a special observation about three dimensions. From our perspective, the fact that the bare action is already finite in 2+1 dimensions is the crucial reason why their construction works.

Now we come to one crucial observation. The renormalized stress-tensor in 2+1 dimensions is given by \cite{balasubramanian}:

\begin{equation}
T^{ren}_{ab} = \frac{1}{\kappa}\left[ \Theta_{ab} - \Theta \gamma_{ab} + \gamma_{ab}\right]
\end{equation}
We will now show that the renormalized Neumann action (which coincidentally happens to be the same as the bare Neumann action in 2+1 dimensions\footnote{This coincidence of the renormalized and the bare Neumann actions is a feature of 2+1 dimensions and does not hold in higher dimensions, but the statements we make about the renormalized action apply in higher dimensions as well.}) gives rise to a well-defined variational principle when we demand that the renormalized boundary stress tensor density is held fixed. This means that, given the renormalized stress-tensor as our boundary data, we have a well defined variational principle. 

To show this, first note that in three dimensions,
\begin{eqnarray}\label{ads3_Nvariation}
\delta S^{ren}_N &=& \delta S_N=\delta \left[ \frac{1}{2\kappa}\int_{\mathcal{M}} d^3 x\;\sqrt{-g}(R-2\Lambda) +\frac{1}{2\kappa}\int_{\partial \mathcal{M}}\;d^2 x\;\sqrt{-\gamma} \Theta \right] \\ \nonumber &=& \frac{1}{2\kappa}\int_{\mathcal{M}} d^3 x\;\sqrt{-g}(G_{ab}+\Lambda g_{ab})\delta g^{ab} - \int d^2 x \left[ \delta \left( -\frac{\sqrt{-\gamma}}{2\kappa}(\Theta^{ab}-\Theta \gamma^{ab}) \right)\gamma_{ab} \right]
\end{eqnarray}
The bare stress-tensor is defined as

\begin{equation}
T^{bare}_{ab}=\frac{1}{\kappa}\left[ \Theta_{ab} - \Theta \gamma_{ab} \right]
\end{equation}
The surface term in \eqref{ads3_Nvariation} can be thus expressed as

\begin{equation}
\delta \left( -\frac{\sqrt{\gamma}}{2\kappa}(\Theta^{ab}-\Theta \gamma^{ab}) \right)\gamma_{ab} = \delta \left(\frac{\sqrt{-\gamma}}{2}T^{bare\;ab} \right)\gamma_{ab}
\end{equation}
Now by an explicit calculation, we can see that

\begin{equation}
\delta \left(\frac{\sqrt{-\gamma}}{2}T^{bare}_{ab} \right)\gamma^{ab} = \delta \left(\frac{\sqrt{-\gamma}}{2}T^{ren}_{ab} \right)\gamma^{ab}\label{ambig}
\end{equation}
This shows that the Neumann variational problem of the renormalized action might as well be formulated by holding the renormalized boundary stress tensor density fixed. This arises because in formulating the variational problem one has the freedom to add a $\chi_{ab}$ to the stress tensor that one is holding fixed at the boundary as long as it satisfies
\bea
\delta\Big(\sqrt{-\gamma} \chi_{ab}\Big) \ \gamma^{ab}=0\label{ambig1}
\eea
We will see that in odd $d$ dimensions, this ambiguity\footnote{We will discuss this ambiguity, together with the logarithmic divergence, elsewhere.} in practice does not arise because the variational problem of Neumann type for the renormalized action essentially automatically leads to the renormalized stress tensor. We turn now to demonstrate this in four dimensions. 

\subsection{AdS$_4$ ($d=3$)}

In $d=3$, the singular part of regularized action evaluates to

\begin{eqnarray}\label{ads4_ren_action}
S_{N}^{reg} &=& -\frac{3}{2\kappa}\int d^3 x \int_{\epsilon} d\rho \frac{\sqrt{-g}}{\rho^{5/2}} \\ \nonumber &=&  -\frac{1}{\kappa}\int d^3 x \sqrt{-g_0} \left( \frac{1}{\epsilon^{3/2}} + \frac{3}{2 \epsilon^{1/2}}\; \Tr\; g_2    \right)
\end{eqnarray}
where we have once again used the determinant expansion \eqref{detexpansion}. The determinant of the induced metric $\gamma_{ab}$ can be expressed as

\begin{equation}\label{h-g_relation}
\sqrt{-\gamma} = \frac{\sqrt{-g}}{\epsilon^{d/2}}
\end{equation}
This, together with \eqref{g2traces} allows us to write the counter-term action

\begin{equation}\label{threed_counterterm}
S^{ct} = \frac{1}{\kappa}\int d^3 x\; \sqrt{-\gamma}\left(1- \frac{1}{4}R[\gamma] \right)
\end{equation}
The fact that this is the correct counter-term can be checked by expanding \eqref{threed_counterterm} in the Fefferman-Graham expansion order by order and using \eqref{detexpansion} and \eqref{g2traces}. The renormalized Neumann action, in a notation analogous to that in \cite{balasubramanian}, is thus given by

\begin{equation}
S^{ren}_N = \frac{1}{2\kappa}\int_{\mathcal{M}} d^4 x \sqrt{-g}\;(R-2\Lambda) + \frac{1}{\kappa}\int d^3 x\; \sqrt{-\gamma}\left(1- \frac{1}{4}R[\gamma] \right)
\end{equation}
Including this counter-term and doing variations, we also reproduce the stress-tensor of \cite{balasubramanian, deHaro}

\begin{equation}
T^{ren}_{ab} = \frac{1}{\kappa}\left[ \Theta_{ab} - \Theta \gamma_{ab} + 2 \gamma_{ab} - G_{ab} \right] \label{4dRenT}
\end{equation}
where $G_{ab}=R_{ab}[\gamma]-\frac{1}{2}R[\gamma] \gamma_{ab}$ is the Einstein tensor of the induced metric\footnote{More precisely, what we reproduce is $\delta T^{ren}_{ab}$ from the variational problem for the renormalized Neumann action. But unlike in odd $d$, this leads directly to (\ref{4dRenT}) and we do not need to use the ambiguity of the type (\ref{ambig1}).}. This stress tensor is known for empty AdS and AdS black hole to be finite and also has the right leading fall-offs to reproduce the correct finite charges for the AdS black hole.

This shows again that the renormalized Neumann action leads to a well-defined variational problem when holding the renormalized boundary stress tensor fixed.

\subsection{AdS$_5$ ($d=4$)}

For the case of $d=4$, the divergent part of the action evaluates to

\begin{eqnarray}\label{ads5_onshellaction}
S_{N}^{reg} &=&  -\frac{2}{\kappa}\int d^4 x \sqrt{-g_0} \left( \frac{3}{2\epsilon^{2}} + \frac{3}{4 \epsilon}\; \Tr\; g_2  - \log \epsilon \;\frac{1}{8} \left( (\Tr (g_2))^2 - \Tr (g_2)^2 \right) \right) \label{4log}
\end{eqnarray}
Barring the log term, all other divergences in \eqref{ads5_onshellaction} can be cancelled by adding a counter-term given by 

\begin{equation}\label{fived_counterterm}
S^{ct}_{N} = \frac{3}{\kappa} \int d^4 x \; \sqrt{-\gamma}
\end{equation}
Once again, this can be explicitly checked by expanding \eqref{fived_counterterm} in Fefferman-Graham expansion and using the relations \eqref{detexpansion} and \eqref{g2traces}. The renormalized Neumann action is given by

\begin{eqnarray}\label{fived_Neumann_action}
S^{ren}_N &=&   \frac{1}{2\kappa}\int_{\mathcal{M}} d^5 x \sqrt{-g}\;(R-2\Lambda) -\frac{1}{2\kappa}\int_{\partial \mathcal{M}}\;d^4 x \sqrt{-\gamma}\; \Theta + \frac{3}{\kappa} \int d^4 x \; \sqrt{-\gamma}
\end{eqnarray}
As in the case of $d=2$, there is an ambiguity in the stress-tensor. The renormalized stress-tensor we hold fixed for the variational principle is given by

\begin{equation}
T^{ren}_{ab} = \frac{1}{\kappa}\left[ \Theta_{ab} - \Theta \gamma_{ab} + 3\gamma_{ab} -\frac{1}{2}G_{ab} \right]
\end{equation}
Once again this shows that the renormalized Neumann action \eqref{fived_Neumann_action} gives a well defined variational principle with renormalized stress-tensor.  We also note that \eqref{fived_Neumann_action}, being an even $d$ case has an ambiguity similar to $d=2$ case, and we have used the fact that 
\bea
\delta \Big(\sqrt{-\gamma}G_{ab}\Big)\ \gamma^{ab}=0.
\eea

In what follows, we will often suppress the superscript ${ren}$ when there is no source of ambiguity that we are indeed working with the renormalized action. 

\subsection{Comparison With Standard Holographic Renormalization}

How does all this compare with the standard discussion of holographic renormalization in the Dirichlet case? 

One difference is that the counter-terms that are added in the Dirichlet case do not change the variational problem: before and after their addition, the boundary metric that is held fixed is identical. This is not true in our case. Before renormalization, the quantity that is held fixed is the {\em unrenormalized} stress tensor density, but at the end it is the {\em renormalized} stress tensor density. It is of course not surprising that added terms can change the variational problem, what is worthy of remark here is the philosophy behind it: we demanded the finiteness of the Neumann action, and that leads to a well-defined variational problem with the {\em renormalized} quantity held fixed. Satisfyingly, this same object can also be obtained as the Legendre transform of the {\em renormalized} Dirichlet action, see Appendix B. Note that the unrenormalized actions are merely a crutch and the renormalized actions are the physically relevant objects.  

Let us also note that the {\em total} action/partition function (including counter-terms {\em and} everything else) can only be a functional of the quantity fixed at the boundary. This is guaranteed at the level of the action because again, the Neumann action is a Legendre transform of Dirichlet and therefore (by construction) depends only on the conjugate variable. In equations, as we discuss in an Appendix, we can view our action as
\begin{equation}\label{Neu_leg_action}
S^{ren}_N [\pi^{ren}_{ab}]= S^{ren}_D[\gamma^{ab}] - \int_{\partial \mathcal{M}}d^{D-1}x\;\pi^{ren}_{ab}\gamma^{ab}
\end{equation}
where 
\bea
\pi^{ren}_{ab} = \frac{\delta S^{ren}_D}{\delta \gamma^{ab}}.
\eea 
This can be viewed as the semi-classical version\footnote{We will briefly discuss the existence of a full quantum theory further in Section 5 and 6, as well as in more detail in \cite{CKCG}.} of a Legendre transform at the level of partition functions:
\bea
\Gamma[\delta W/\delta \gamma^{ab}]=W[\gamma^{ab}]- \int_{\partial \mathcal{M}}d^{D-1}x\; \frac{\delta W}{\delta \gamma^{ab}}\gamma^{ab}
\eea
At the level of the semi-classical saddle, this translates to the statement that the variational principle (while holding the conjugate quantity fixed at boundary) is well-defined, which we checked explicitly earlier in this section. The separate terms (including counter-terms) in the action which are  integrated over can have complicated dependences, but they conspire to satisfy the above demands. 

As an aside, we also note some papers in the literature which deal with related set-ups. In particular, in \cite{Geoffrey} the boundary metric fluctuates but they arrange that the variational principle with the Dirichlet action works, by setting $T^{ij}=0$.  There are other papers, especially in three dimensions, which deal with similar set-ups  \cite{Apolo, cases, Myers}. In fact, our approach can be thought of in many ways as a general framework for dealing with some of these situations. The work of \cite{Geoffrey} treats the boundary stress tensor to be a fixed given {\em value} (namely, zero), so their partition function is a {\em number}, so they do not discuss the points we emphasize in the previous paragraph. Our work can be thought of as a generalization of theirs and our partition function is a proper functional, where instead of setting the stress tensor (density) to be zero, we treat it as arbitrary but fixed\footnote{The ``arbitrariness" of the boundary stress tensor should of course still satisfy the requirement that the Fefferman-Graham expansion should satisfy the bulk equations of motion, see the discussion in \cite{deHaro} for details.}.

\section{Finite On-shell Action}

In this section we present the results of on-shell action and stress-energy tensor for the Neumann action in various dimensions. We also draw comparison of our on-shell action with the on-shell Dirichlet action. Note that the precise value of the action is sensitive to the infrared cutoff of the action integral. So one cannot work abstractly at the level of the Fefferman-Graham expansion like we did so far, because we need to know the metric finitely deep into the geometry and not merely as an expansion at the boundary. So we will consider explicit solutions like black holes.

\subsection{AdS$_3$}

The Dirichlet action for gravity in AdS$_3$ is given by \cite{balasubramanian}

\begin{equation}
S_D = \frac{1}{2\kappa}\int_{\mathcal{M}}d^3x\;\sqrt{-g}(R-2\Lambda)+ \frac{1}{\kappa}\int_{\partial\mathcal{M}}\sqrt{-\gamma}\Theta -\frac{1}{\kappa}\int_{\partial\mathcal{M}}\sqrt{-\gamma}
\end{equation}
We evaluate the above action on the BTZ metric

\begin{equation}
ds^2 = -\frac{(r^2 - r^{2}_{+})(r^2 - r^{2}_{-})}{r^2}dt^2 + \frac{r^2\;dr^2}{(r^2 - r^{2}_{+})(r^2 - r^{2}_{-})} + r^2 \left(d\phi - \frac{r_{+}r_{-}}{r^2}dt \right)^2
\end{equation}
where $r_{+}$ and $r_{-}$ are the outer and inner horizons respectively and are related to the charges through the relation $M=r_{+}^{2}+r_{-}^{2}$ and $J=2r_{+}r_{-}$. In the above metric we have set $l=1$. Evaluating the action between time $-T$ to $T$ and $r_{+}<r<R$ on this solution yields

\begin{equation}
S_{D}^{BTZ} = \frac{2\pi(r_{+}^2+r_{-}^2)T}{\kappa} + O\left(\frac{1}{R^2}\right)
\end{equation}
The on-shell Neumann action for the BTZ solution yields

\begin{equation}
S_{N}^{BTZ} = \frac{2\pi(r_{+}^2+r_{-}^2)T}{\kappa}
\end{equation}
which matches with the Dirichlet action in the limit $R\rightarrow \infty$. The stress-energy tensor similarly takes the form

\begin{equation}
T_{ab} = \left( \begin{array}{cc}
-\frac{r_{+}^2+r_{-}^2}{2\kappa}  & \frac{r_{+}r_{-}}{\kappa}  \\
\frac{r_{+}r_{-}}{\kappa}  & -\frac{r_{+}^2+r_{-}^2}{2\kappa} 
\end{array} \right)+ O\left(\frac{1}{R^2}\right)
\end{equation}
This stress tensor has the right fall-offs to reproduce finite charges $M$ and $J$ through the relation \cite{brown, quasilocal}

\begin{equation}
Q_{\xi} = -\int_{\Sigma}d^{D-1}x\;\sqrt{\sigma}(u^{a}T_{ab}\xi^{b})
\end{equation}
where $\xi^{a}$ is the Killing vector generating the isometry of the boundary metric and $u^{a}$ is the unit time-like vector. We see that the counter-term action that was chosen to make the on-shell Neumann action finite also produces a finite stress tensor.  This was shown for the Dirichlet case by \cite{balasubramanian}.

\subsection{AdS$_4$}

The (renormalized) Dirichlet action in $D=4$ takes the form

\begin{equation}
S_D = \frac{1}{2\kappa}\int_{\mathcal{M}}d^4x\;\sqrt{-g}(R-2\Lambda)+ \frac{1}{\kappa}\int_{\partial\mathcal{M}}d^3x\;\sqrt{-\gamma}\Theta -\frac{2}{\kappa}\int_{\partial\mathcal{M}}d^3x\;\sqrt{-\gamma}\left(1+\frac{{}^{(3)}R}{4}\right)
\end{equation}
The AdS-Schwarzschild black hole metric is given by

\begin{equation}
ds^2 = -(1-\frac{2M}{r}+r^2)dt^2 + \frac{dr^2}{(1-\frac{2M}{r}+r^2)} + r^2d\Omega^2
\end{equation}
The horizon is obtained by the real root of

\begin{equation}
1-\frac{2M}{r_H}+r_{H}^2 = 0
\end{equation}
Evaluating the action for this metric yields (integrated in the region $-T<t<T$ and $r_{H}<r<R$)

\begin{equation}
S_{D}^{AdS-BH} = -\frac{8\pi (M-r_{H}^3) T}{\kappa} + O\left(\frac{1}{R}\right)
\end{equation}
The stress tensor computed for this metric is given by

\begin{equation}
T_{ab}=\left(
\begin{array}{ccc}
 -\frac{2 M }{\kappa R } & 0 & 0 \\
 0 & -\frac{M}{\kappa R } & 0 \\
 0 & 0 & -\frac{M \sin ^2(\theta )}{\kappa R} \\
\end{array}
\right)+O\left(1/R^2 \right)
\end{equation}
which once again has the right fall-offs to obtain finite charges as described in the previous section. The Neumann action in $D=4$ takes the form

\begin{equation}
S_{N} = \frac{1}{2\kappa}\int_{\mathcal{M}}d^4x\;\sqrt{-g}(R-2\Lambda)+\frac{1}{\kappa}\int_{\partial\mathcal{M}}d^3x\;\sqrt{-\gamma}\left(1-\frac{{}^{(3)}R}{4} \right)
\end{equation}
which evaluates to

\begin{equation}
S_{N}^{AdS-BH} = -\frac{8\pi (M-r_{H}^3) T}{\kappa} + O\left(\frac{1}{R}\right)
\end{equation}
The sub-leading term here differs from the sub-leading term in the Dirichlet action and the two actions are same only in the $R\rightarrow \infty$ limit.

\subsection{AdS$_5$}

In $D=5$ the Dirichlet action takes the form

\begin{equation}
S_D = \frac{1}{2\kappa}\int_{\mathcal{M}}d^5x\;\sqrt{-g}(R-2\Lambda)+ \frac{1}{\kappa}\int_{\partial\mathcal{M}}d^4x\;\sqrt{-\gamma}\Theta -\frac{3}{\kappa}\int_{\partial\mathcal{M}}d^3x\;\sqrt{-\gamma}\left(1+\frac{{}^{(4)}R}{12}\right)
\end{equation}
Evaluating this action for the black hole metric

\begin{equation}
ds^2 = -f(r)dt^2 + \frac{dr^2}{f(r)}+r^2d\Omega_{3}^{2}
\end{equation}
where

\begin{equation}
f(r) = r^2 + 1 -\frac{2M}{r^2}
\end{equation}
The horizon is once again determined by the largest positive root of

\begin{equation}
r_{H}^2 + 1 -\frac{2M}{r_{H}^2} = 0
\end{equation} 
The action evaluates to
\begin{equation}
S_{D}^{BH} = -\frac{2\pi^2T}{\kappa}(2M+\frac{3}{4}-2r_{H}^{4}) + O\left(1/R^4 \right)
\end{equation}
The stress tensor takes the form

\begin{equation}
T_{ab} = \left(
\begin{array}{cccc}
 -\frac{3 (8M+1) }{8R^2 \kappa } & 0 & 0 & 0 \\
 0 & -\frac{(8M+1)}{8R^2 \kappa } & 0 & 0 \\
 0 & 0 & -\frac{\left((8M+1) \sin ^2(\psi )\right)}{8R^2
   \kappa } & 0 \\
 0 & 0 & 0 & -\frac{\left((8M+1) \sin ^2(\theta ) \sin
   ^2(\psi )\right) }{8R^2 \kappa } \\
\end{array}
\right)+O\left(1/R^4\right)
\end{equation}
The Neumann action in this case can be written as

\begin{equation}
S_N = \frac{1}{2\kappa}\int_{\mathcal{M}}d^5x\;\sqrt{-g}(R-2\Lambda)- \frac{1}{\kappa}\int_{\partial\mathcal{M}}d^4x\;\sqrt{-\gamma}\Theta +\frac{3}{\kappa}\int_{\partial\mathcal{M}}d^4x\;\sqrt{-\gamma}
\end{equation}
which evaluates to

\begin{equation}
S_{N}^{BH} = -\frac{2\pi^2T}{\kappa}(2M+\frac{3}{4}-2r_{H}^{4}) + O\left(1/R^2 \right)
\end{equation}
Again, we find agreement when the radial cutoff is taken to infinity.

\section{ADM Formulation of Renormalized AdS$_4$ Action}

The ADM formulation of GR works by singling out the time direction from the spatial direction and re-expressing the content of GR in terms of ADM variables. Thus the spacetime is thought of as foliated by spatial slices $\Sigma_t$ which are the hypersurfaces of constant $t$. The spacetime metric can be expressed as

\begin{equation}\label{adm_metric_full}
ds^2 \equiv g_{\alpha \beta}dx^{\alpha} dx^{\beta} = -N^2 dt^2 + h_{ab}(dy^a + N^{a}dt)(dy^b + N^{b}dt) 
\end{equation}
where $N$ is the Lapse function, $N^{a}$ is the shift vector and $h_{ab}$ is the induced metric on the hypersurface $\Sigma_t$. In what follows, we assume that the manifold is a box with finite spatial extend such that the boundary is time-like, denoted ${\cal B}$. The spatial section of ${\cal B}$ is denoted $B$. We will also ignore the space-like boundaries at initial and final times and work with coordinates such that the time-like boundary is orthogonal to the spatial hypersurfaces, $\Sigma_t$. Under the ADM split of the bulk metric, \eqref{adm_metric_full}, the induced metric on the boundary ${\cal B}$, also undergoes a decomposition

\begin{equation}\label{gamma_ADM}
ds^2 \equiv \gamma_{ij}dx^{i} dx^{j} = -N^2 dt^2 + \sigma_{AB}(d\theta^A + N^{A}dt)(d\theta^{B} + N^{B}dt)
\end{equation} 
where $\sigma_{AB}$ is the induced metric on $B$. We will also need the expression for the decomposition of Ricci scalar

\begin{equation}\label{ricci_decomp}
{}^{(D)} R = {}^{(D-1)}R + K^{ab}K_{ab} - K^2 - 2\nabla_{\alpha}\left( u^{\beta} \nabla_{\beta}u^{\alpha} - u^{\alpha} \nabla_{\beta} u^{\beta} \right)
\end{equation}
where $K_{ab}$ is the extrinsic curvature of the spatial hypersurface $\Sigma_t$ (not to be confused with the boundary). The point about ADM split is that $N$ and $N^a$ are not dynamical fields and therefore their conjugates are constraint relations. The dynamical field is the spatial metric $h_{ab}$ and the canonical conjugate momentum is given by

\begin{equation}
p^{ab} \equiv \frac{\partial}{\partial \dot{h}_{ab}}(\sqrt{-g}\mathcal{L}_G) = \frac{\sqrt{h}}{2\kappa}(K^{ab}-Kh^{ab})
\end{equation}
where $K_{ab}$ is the extrinsic curvature of $\Sigma_t$. The details of the ADM decomposition of gravitational action can be found in \cite{poisson, latest}. We will work with AdS$_4$ in what follows, for convenience.

\subsection{Dirichlet action}
In this section, we return to the case of ADM decomposition, for the renormalized Dirichlet action in AdS$_4$. The renormalized action in the covariant form is given by \cite{deHaro,balasubramanian}

\begin{eqnarray}\label{D_ads_cov_action}
S_D &=& \frac{1}{2\kappa}\int_{\mathcal{M}}d^4x\;\sqrt{-g}(R-2\Lambda)+ \frac{1}{\kappa}\int_{\cal B}d^3x\;\sqrt{-\gamma}\Theta \\ \nonumber &+& \frac{1}{\kappa}\int_{\cal B}d^3x\;\sqrt{-\gamma}\left(-\frac{2}{l} \right) \left[1+ \frac{{}^{(3)}R}{4} \right]
\end{eqnarray}
The first two terms are the Einstein-Hilbert and the GHY piece, and can be written in terms of the ADM variables following the steps of \cite{latest}. This gives us the following form for the action \cite{poisson}

\begin{eqnarray}
S_D &=& S_{EH} + S_{GHY} + S_{ct} \\ \nonumber &=& \int_{\cal M}d^D x \left( p^{ab}\dot{h}_{ab} - NH - N_a H^a \right) +  \int_{\cal B}d^{D-1}y \sqrt{\sigma}\left(N\varepsilon - N^a j_a \right) + S_{ct}
\end{eqnarray}
where $H$ and $H^a$ are the Hamiltonian and momentum constraints,

\begin{eqnarray}
H &=& \frac{\sqrt{h}}{2\kappa}\left( K^{ab}K_{ab} - K^2 - {}^{(3)}R + 2\Lambda \right) \\ \nonumber
H^a &=& -\frac{\sqrt{h}}{\kappa}D_b(K^{ab}-Kh^{ab}) \\ \nonumber
\end{eqnarray}
$\sqrt{\sigma}\varepsilon$, $\sqrt{\sigma}j_a$ and $N\sqrt{\sigma}s^{ab}/2$ are the momenta conjugate to $N$, $N^{a}$ and $\sigma_{ab}$. and are given by

\begin{eqnarray}\label{canonical_variables}
\varepsilon &=& \frac{k}{\kappa}, \quad  j_a = \frac{2}{\sqrt{h}}r_b p_{\;a}^{b} \\ \nonumber s^{ab} &=& \frac{1}{\kappa}\left[k^{ab} - \left(\frac{r^a \partial_a N}{N} + k\right)\sigma^{ab} \right]
\end{eqnarray} 
where $k^{ab}$ is the extrinsic curvature of $B$ embedded in $\Sigma_t$ and $k=k^{ab}\sigma_{ab}$. The counter-term action is given in the covariant form by

\begin{equation}
S_{ct} = \frac{1}{\kappa}\int_{\cal B}d^3x\;\sqrt{-\gamma}\left(-\frac{2}{l} \right) \left[1+ \frac{{}^{(3)}R}{4} \right]
\end{equation}
Using \eqref{ricci_decomp} and the expression for the determinant $\sqrt{-\gamma}=N\sqrt{\sigma}$, we obtain the counter-term action as

\begin{equation}
S_{ct} = \frac{1}{\kappa}\int_{\cal B}d^3x\left(-\frac{2}{l} \right)\left[1 + \frac{l^2}{4}\left( {}^{(2)}R + \hat{k}_{ab}\hat{k}^{ab} - \hat{k}^2 \right) \right]
\end{equation}
where $\hat{k}_{ab}$ is the extrinsic curvature of $B$ as a hypersurface embedded in ${\cal B}$. For black hole geometries, we also get a contribution from the horizon which is given by decomposing the covariant Neumann action with a boundary at the horizon where no data is specified \cite{martinez, brown, latest}. The action then takes the form

\begin{eqnarray}
S_D &=& \int_{\cal M}d^D x \left( p^{ab}\dot{h}_{ab} - NH - N_a H^a \right) \\ \nonumber &+& \int_{\cal H}d^{D-1}y\;\sqrt{\sigma}\left(\frac{r^a \partial_a N}{\kappa} + \frac{2r_a N_b p^{ab}}{\sqrt{h}} \right) + \int_{\cal B}d^{D-1}y \sqrt{\sigma}\left(N\varepsilon - N^a j_a \right) \\ \nonumber &+& \frac{1}{\kappa}\int_{\cal B}d^3x\left(-\frac{2}{l} \right)\left[1 + \frac{l^2}{4}\left( {}^{(2)}R + \hat{k}_{ab}\hat{k}^{ab} - \hat{k}^2 \right) \right]
\end{eqnarray}
We can further express the above action in terms of the renormalized parameters thereby absorbing the counter-term into the renormalized quantities $\varepsilon^{ren}=\varepsilon+\varepsilon^{ct}$, $j_{a}^{ren}=j_{a}+j^{ct}_{a}$ and $s_{ab}^{ren}=s_{ab}+s_{ab}^{ct}$. To do so, we do a canonical decomposition of the tensor using normal and tangential projections \cite{quasilocal}. The expressions for renormalized quantities are given by

\begin{eqnarray}
\varepsilon^{ren} &=& u_a u_b T^{ab} \\ \nonumber
j_{a}^{ren} &=& -\sigma_{ab}T^{bc}u_c \\ \nonumber
s_{ab}^{ren} &=& \sigma_{ac}\sigma_{bd} T^{cd} \nonumber
\end{eqnarray} 
where $T^{ab}$ is the renormalized stress tensor given by

\begin{equation}
T^{ab} = \frac{1}{\kappa}\left( \Theta^{ab}-\Theta \gamma^{ab} + \frac{2}{l}\gamma^{ab} -l G^{ab} \right)
\end{equation}
Using the above expressions, we get

\begin{eqnarray}
\varepsilon^{ren} &=& \varepsilon - \frac{1}{\kappa}\left[ \frac{2}{l} + \frac{l}{2}\left({}^{(2)}R - \hat{k}_{ab}\hat{k}^{ab} + \hat{k}^2 \right) \right] \\ \nonumber
j_{a}^{ren} &=& j_a + \frac{l}{\kappa}\left(d_{a}\hat{k} - d_{b}\hat{k}^{b}_{\; a} \right) \\ \nonumber
s_{ab}^{ren} &=& s_{ab} + \frac{1}{\kappa}\left[ \frac{2}{l}\sigma_{ab} + \frac{l}{2}\left( {}^{(2)}R + \hat{k}^{ab}\hat{k}_{ab} - \hat{k}^2 \right) \right. \\ \nonumber &-& \left. l\left( -\frac{1}{N}{\cal L}_{m}\hat{k}_{ab}- \frac{1}{N}d_a d_b N + {}^{(2)}R_{ab}+ \hat{k}\hat{k}_{ab}-2 \hat{k}_{ac}\hat{k}^{c}_{b} \right) \right]
\end{eqnarray}
In writing the above expressions, we have made use of Gauss-Codazzi relations whose exact expressions are given in the Appendix. Thus, the renormalized action can be expressed as

\begin{eqnarray}\label{D_ads_adm_action}
S_D &=& \int_{\cal M}d^D x \left( p^{ab}\dot{h}_{ab} - NH - N_a H^a \right) \\ \nonumber &+& \int_{\cal H}d^{D-1}y\;\sqrt{\sigma}\left(\frac{r^a \partial_a N}{\kappa} + \frac{2r_a N_b p^{ab}}{\sqrt{h}} \right) + \int_{\cal B}d^{D-1}y \sqrt{\sigma}\left(N\varepsilon^{ren} - N^a j_{a}^{ren} \right)
\end{eqnarray}

\subsubsection{Kerr-AdS: Covariant}

As an illustration of our construction, we can evaluate the action on the Kerr-AdS metric in $D=4$. Rotating black holes are better defined in AdS, than flat space (see eg., \cite{CKRotation, CKTomogram}). The metric in Boyer-Lindquist type coordinates is given by

\begin{eqnarray}\label{kerr_metric}
ds^2 = \rho^2\left( \frac{dr^2}{\Delta}+\frac{d\theta^2}{\Delta_{\theta}}\right)+\frac{\Delta_{\theta}\sin^2\theta}{\rho^2}\left(adt - \frac{r^2 + a^2}{\Sigma}d\phi \right)^2 - \frac{\Delta}{\rho^2}\left( dt - \frac{a\sin^2 \theta}{\Sigma}d\phi \right)^2
\end{eqnarray}
where

\begin{eqnarray} 
\rho^2 &=& r^2 + a^2\cos^2 \theta, \qquad \Delta = (r^2 + a^2)\left(1+\frac{r^2}{l^2} \right)-2Mr \\ \nonumber \Delta_{\theta} &=& 1 - \frac{a^2}{l^2}\cos^2 \theta, \qquad \Sigma = 1- \frac{a^2}{l^2}
\end{eqnarray}
The horizon is at the largest positive root of $\Delta(r_H)=0$. The angular velocity of the black hole (for $r \geq r_H$) is given by
\begin{equation}
\omega = a\Sigma \left( \frac{\Delta_{\theta}(r^2 + a^2)- \Delta}{(r^2 + a^2)^2 \Delta_{\theta} - a^2 \Delta \sin^2 \theta} \right)
\end{equation}
The angular velocity at the horizon is given by

\begin{equation}
\Omega_H  = \frac{a\Sigma}{r_H^{2}+a^2}
\end{equation}
while the angular velocity at the boundary ($r \rightarrow \infty$), is given by $\Omega_{\infty} = -a/l^2$. The angular velocity relevant for the thermodynamics is given by $\Omega = \Omega_H - \Omega_{\infty}$ \cite{Klemm, HHT}.
Given the metric, the ADM variables can be read off by comparing \eqref{kerr_metric} with the ADM form of the metric. The Lapse, Shift and spatial metric is given by
\begin{eqnarray}
N &=& \sqrt{\frac{\rho^2 \Delta \Delta_{\theta}}{(r^2 + a^2)^2 \Delta_{\theta}-a^2 \Delta \sin^2 \theta}} \\ \nonumber
N^{\phi} &=& a\Sigma \frac{\left( \Delta - \Delta_{\theta}(r^2 + a^2) \right)}{(r^2 + a^2)^2 \Delta_{\theta}-a^2 \Delta \sin^2 \theta} \\ \nonumber
h_{ab} &=& \left( \begin{array}{ccc}
\frac{\rho^2}{\Delta} & 0 & 0 \\
0 & \frac{\rho^2}{\Delta_\theta} & 0 \\
0 & 0 & \frac{\left( (r^2 + a^2)^2 \Delta_\theta - a^2 \Delta \sin^2 \theta  \right)}{\rho^2 \Sigma^2}
\end{array} \right) \nonumber
\end{eqnarray}

For thermodynamic interpretation we must work with the complex metric associated with the black hole, which is given by the identification $N \rightarrow -i\tilde{N}, N^{\phi} \rightarrow -i\tilde{N}^{\phi}$ \cite{brown, latest}. The periodicity of time circle can be estimated by evaluating $r^a\partial_a \tilde{N} \equiv 2\pi/\beta$ term on the horizon. This gives the time periodicity, $\beta$, to be

\begin{equation}\label{beta}
\beta = \frac{4\pi (r^{2}_{H} + a^2)}{r_H\left(1+ \frac{a^2}{l^2} + \frac{3r_{H}^2}{l^2}-\frac{a^2}{r_{H}^2}\right)}
\end{equation}
The expressions for various terms in the covariant action are:

\begin{eqnarray}
R &=& -\frac{12}{l^2} \\ \nonumber
\Theta &=& \frac{3}{l} + \frac{(-3a^2 + 2l^2 -5a^2 \cos 2\theta)}{4l R_c^2} + O(1/R_c^4) \\ \nonumber
{}^{(3)}R &=& \frac{2l^2 -3a^2 -5a^2 \cos 2\theta}{l^2 R_c^2} + O(1/R_c^4)
\end{eqnarray}
Evaluating the complex metric on the covariant action \eqref{D_ads_cov_action}, and using the expression \eqref{beta} for the periodicity, we get

\begin{equation}
S_D = -i \frac{\pi l^2 (r_{H}^{2}+a^2)^2 (l^2 - r_{H}^{2})}{(l^2 - a^2)\left(a^2 l^2-(a^2 + l^2)r_{H}^2 - 3r_{H}^4 \right)}
\end{equation}
This is related to the free energy through the relation

\begin{equation}\label{D_ads_free_energy}
-\beta F_D \equiv \log Z_D \approx iS_D
\end{equation}
where $\beta$ is the inverse temperature which can be identified with the periodicity of the time circle. This gives the free energy of the black hole to be

\begin{equation}\label{cov_Dir_FE}
F_D = \frac{(r_{H}^2 + a^2)(l^2 - r_{H}^{2})}{4(l^2-a^2)r_H}
\end{equation}

\subsubsection{Kerr-AdS: ADM}
Evaluating the complex metric on the ADM decomposed action, the bulk term vanishes because the metric is stationary and satisfies Einstein's equation. The horizon term gives a contribution of

\begin{equation}
S_{\cal H} = -i \frac{A}{4} - i\Omega_H P J
\end{equation}
On the boundary we can see that the renormalized $\varepsilon$, $j^\phi$ and $s^{AB}$ have correct fall-offs so as to give finite results for the integral,

\begin{eqnarray}
\varepsilon_{ren} &=& \left( \frac{M(a^2 - 4l^2 + 3a^2 \cos 2\theta)}{l\Sigma \kappa}  \right)\frac{1}{R_c^3} + O(\frac{1}{R_c^4}) \\ \nonumber
j^{\phi}_{ren} &=& \frac{3aM}{\kappa}\sqrt{\frac{\Delta_{\theta}}{\Sigma}}\frac{1}{R_c^4} + O(\frac{1}{R_c^5}) \\ \nonumber
\end{eqnarray}
Evaluating the boundary integrals, we have
\begin{equation}
S_{\cal B} = iEP + i \Omega_{\infty}P J
\end{equation}
where $E$ and $J$ are calculated as

\begin{equation}
E = \frac{M}{\Sigma^2},\qquad J = \frac{Ma}{\Sigma^2}
\end{equation}
which are the ADM charges of the Kerr black hole. Using \eqref{D_ads_free_energy}, we have

\begin{equation}\label{adm_Dir_FE}
F_D = E - TS -\Omega J
\end{equation}
Now, by an explicit computation, we can verified that the free energy, $F_D$, in \eqref{cov_Dir_FE} can be expressed as

\begin{equation}\label{Dir_cov_FE_new}
F_D = -T \frac{A}{4} -\Omega J + g(A,J)
\end{equation}
where
\begin{equation}
g(A,J) = \sqrt{\frac{A}{16\pi}+\frac{4\pi}{A}J^2 + \frac{J^2}{l^2} + \frac{A}{8\pi l^2}\left(\frac{A}{4\pi} + \frac{A^2}{32\pi^2 l^2} \right)}
\end{equation}
Equating \eqref{Dir_cov_FE_new} to the free energy computed using ADM approach, \eqref{adm_Dir_FE}, we get the generalized Smarr formula (see eq.(41) of \cite{Klemm}).

\begin{equation}
E^2 = \frac{A}{16\pi}+\frac{4\pi}{A}J^2 + \frac{J^2}{l^2} + \frac{A}{8\pi l^2}\left(\frac{A}{4\pi} + \frac{A^2}{32\pi^2 l^2} \right)
\end{equation}
Following \cite{Klemm} we can also relate these calculations to the first law, which we will not repeat. 
 
\subsection{Neumann action}

The renormalized Neumann action in AdS$_4$ is given by

\begin{equation}
S_N = \frac{1}{2\kappa}\int_{\mathcal{M}}d^4x\;\sqrt{-g}(R-2\Lambda) + \frac{1}{\kappa}\int_{\cal B}d^3x \sqrt{-\gamma}\left(\frac{1}{l}\right) \left[1-\frac{l^2}{4}{}^{(3)}R \right]
\end{equation}
The bare part of the Neumann action in ADM was derived in \cite{latest}. In $D=4$ it can be used to write

\begin{eqnarray}
S_N &=& S_{EH} + S_{ct} \\ \nonumber &=& \int_{\mathcal{M}}d^4 x \left( p^{ab}\dot{h}_{ab} - NH - N_a H^a \right) +  \int_{\mathcal{B}}d^{3}x \sqrt{\sigma}\left(\frac{N\varepsilon}{2}  - N^a j_a + \frac{N}{2}s^{ab}\sigma_{ab} \right) + S_{ct}
\end{eqnarray}
The counter-term action can be decomposed similar to the Dirichlet case and we get

\begin{equation}
S_{ct} = \frac{1}{\kappa}\int_{\cal B}d^3x\left(\frac{1}{l} \right)\left[1 - \frac{l^2}{4}\left( {}^{(2)}R + \hat{k}_{ab}\hat{k}^{ab} - \hat{k}^2 \right) \right]
\end{equation}
For the black hole geometries, one again has a contribution from the horizon and action takes the form

\begin{eqnarray}
S_N &=& \int_{\cal M}d^4 x \left( p^{ab}\dot{h}_{ab} - NH - N_a H^a \right) \\ \nonumber &+&\int_{\mathcal{H}}d^{3}y\;\sqrt{\sigma}\left(\frac{r^a \partial_a N}{\kappa} + \frac{2r_a N_b p^{ab}}{\sqrt{h}} \right) + \int_{\mathcal{B}}d^{3}x \sqrt{\sigma}\left(\frac{N\varepsilon}{2}  - N^a j_a + \frac{N}{2}s^{ab}\sigma_{ab} \right) \\ \nonumber &+& \frac{1}{\kappa}\int_{\cal B}d^3x\left(\frac{1}{l} \right)\left[1 - \frac{l^2}{4}\left( {}^{(2)}R + \hat{k}_{ab}\hat{k}^{ab} - \hat{k}^2 \right) \right]
\end{eqnarray}
Using the expressions for renormalized parameters, the Neumann action can be expressed as

\begin{eqnarray}
S_N &=& \int_{\cal M}d^4 x \left( p^{ab}\dot{h}_{ab} - NH - N_a H^a \right) \\ \nonumber &+&\int_{\mathcal{H}}d^{3}y\;\sqrt{\sigma}\left(\frac{r^a \partial_a N}{\kappa} + \frac{2r_a N_b p^{ab}}{\sqrt{h}} \right) + \int_{\mathcal{B}}d^{3}x \sqrt{\sigma}\left(\frac{N\varepsilon^{ren}}{2}  - N^a j^{ren}_a + \frac{N}{2}s^{ren\;ab}\sigma_{ab} \right)
\end{eqnarray}

\subsubsection{Kerr-AdS: Covariant}
We can evaluate the covariant Neumann action on the Kerr-AdS complex metric, we obtain

\begin{equation}
S_N = -i \frac{\pi l^2 (r_{H}^{2}+a^2)^2 (l^2 - r_{H}^{2})}{(l^2 - a^2)\left(a^2 l^2-(a^2 + l^2)r_{H}^2 - 3r_{H}^4 \right)}
\end{equation}
Notice that unlike the asymptotically flat case, the on-shell value of the Dirichlet and Neumann action are equal. The on-shell action is related to the Neumann free energy through the relation

\begin{equation}\label{N_ads_free_energy}
-\beta F_N \equiv \log Z_N \approx iS_N
\end{equation}
which gives the free energy of the black hole to be

\begin{equation}
F_N = \frac{(r_{H}^2 + a^2)(l^2 - r_{H}^{2})}{4(l^2-a^2)r_H}
\end{equation}

\subsubsection{Kerr-AdS: ADM}
Evaluating the complex metric on the ADM decomposed action, the horizon term gives a contribution of

\begin{equation}
S_{\cal H} = -i \frac{A}{4} - i\Omega_{H}P J
\end{equation}
On the boundary we have,

\begin{eqnarray}
s^{ab}_{ren} &=& \left( \begin{array}{cc}
-\frac{lM\Delta_\theta}{\kappa} & 0 \\
0 & -\frac{M(a^2 + 2l^2 -3a^2\cos 2\theta)}{2l\kappa \sin^2\theta}
\end{array} \right)\frac{1}{R_c^3} + O(1/R_c^4) \\ \nonumber
\sigma_{ab} &=& \left( \begin{array}{cc}
\frac{\rho^2}{\Delta_\theta} & 0 \\
0 & \frac{\left( (r^2 + a^2)^2 \Delta_\theta - a^2 \Delta \sin^2 \theta  \right)}{\rho^2 \Sigma^2}
\end{array} \right) \nonumber
\end{eqnarray}
We get a contribution of $iEP/2$ from the integration over $\varepsilon^{ren}$ term and another contribution of $iEP/2$ from the integration over $s_{ab}^{ren}$ term. The $j^{ren}_{\phi}$ gives a contribution of $i\Omega_{\infty} P J$. Together we have again
\begin{equation}
S_{\cal B} = iEP - i\Omega_{\infty}P J
\end{equation}
Again using \eqref{N_ads_free_energy}, the free energy takes the form

\begin{equation}
F_N = E - TS - \Omega J
\end{equation}
where $\Omega = \Omega_H - \Omega_{\infty}$ is the potential relevant for the thermodynamics. So we end up getting the exact same expressions for $F_N$ and $F_D$ (in covariant and canonical approaches, separately).

The emergence of the canonical ensemble together with the Smarr formula implies the first law as well. This follows from the discussion in \cite{Klemm}, so we will not repeat it. 

\section{Alternative Quantizations in AdS}

We would like to investigate whether these boundary conditions can define a consistent quantum gravity in AdS. If so, this will provide a set of boundary conditions that are different from the standard Dirichlet boundary conditions familiar from AdS/CFT.  So far on the other hand, a skeptic could choose to think of our discussion as merely a class of well-defined boundary conditions/terms for {\em classical} gravity in AdS. However, the fact that these boundary conditions give rise to finite actions that lead to correct thermodynamical relations is suggestive to us of an underlying quantum theory: so let us try and explore to see whether we can take these boundary conditions seriously at the quantum level\footnote{CK thanks K. Skenderis for comments on (non-)normalizable modes and the choice of quantizations in AdS.}. We will not prove  in this paper (but see \cite{CKCG}) that our approach can be the starting point of a consistent quantum theory, but we will merely make some related observations. 

From the boundary theory point of view, the translation from the metric-fixed to stress-tensor-fixed point of view is a Legendre transform that takes the boundary partition function to the boundary effective action\footnote{A similar approach for scalar fields was taken in \cite{KW}, the source and condensate are dual variables in the Legendre transform sense.}. This seems to us to be a perfectly natural and consistent operation as we discussed in Section 2.5, so we believe there should be a legitimate formulation of holography in which the correspondence is phrased in the language of the effective action and not in terms of the generating functional. Note that for this, we will have to move away from the standard Dirichlet formulation of holography where the boundary values of bulk fields are interpreted as sources.

The trouble is that it is well-known that (for example) for scalars in a fixed AdS background, of the two modes (which we can call Dirichlet and Neumann) only the Dirichlet mode is typically normalizable \cite{KW, Brietenlohner}. The exception to this is when the mass of the scalar falls in the Brietenlohner-Freedman window, where a Legendre transform analogous to ours takes the Dirichlet scalar theory to the Neumann scalar theory, and both are well-defined quantum mechanically \cite{KW}. When the scalar mass is not in this specific range, there is only one choice of acceptable normalizable mode and a unique quantization in a fixed AdS background.

To understand this better, let us note that the reason why we want normalizable modes is because we want them to be well-defined states in the Hilbert space of the putative quantum theory, with finite norm. This translates to a notion of finite energy: when the scalar mode has finite energy in the bulk of AdS, it can be well-defined as a state in the Hilbert space of the quantum theory. This is what happens in the case of scalar quantum field theory in a fixed AdS background \cite{Isham, Brietenlohner, KW}. 

Now, lets consider the case when the background is not rigid and the metric is allowed to fluctuate. Lets start by considering scalar fields in such a set up. We note two things. One is that a dynamical background makes the notion of energy more subtle, and secondly the notion of mass of the scalar is ambiguous because (say) a term of the form
\bea
(m^2+\lambda R) \phi^2 
\eea
where $R$ is the curvature scalar of the background will look like a usual mass term in the rigid limit. So a non-minimal coupling can sometimes be difficult to distinguish. As it happens both these issues have been addressed in \cite{MincesRivelles} (see also \cite{MincesScalar, RivellesSummary}) and it was found that once one deals with the appropriate notion of (canonical) energy both quantizations are admissible. We will take this as an encouraging fact: when dealing with the full gravity theory with appropriate counterterms etc. it is not necessarily only a Dirichlet boundary condition that can be well-defined, the notion of canonical energy needs to take into account the full theory.

Indeed, a similar conclusion was arrived at by Compere and Marolf \cite{Geoffrey}, who considered the possibility of not fixing the boundary metric, and instead considered simply integrating it over in the path integral. At the semi-classical level, the variational principle would then yield 
\begin{equation}
 \delta S^{ren}_{D} = \text{ Eqs. of motion} + \frac{1}{2} \int_{\partial M} d^d x \sqrt{-g_{0}}
 T^{ij} \delta g_{0_{ij}},
 \end{equation}
where now there is no assumption that $\delta g_{0_{ij}}=0$ because we are letting it fluctuate. This means that to ensure that the action is stationary, now we need the boundary (renormalized) stress tensor to vanish\footnote{The boundary stress-energy tensor that we have often used in our discussions in this paper is given by the relation 
\begin{equation}
T^{ren}_{ij}[\gamma] = -\frac{2}{\sqrt{-\gamma}}\frac{\delta S^{ren}_D}{\delta \gamma^{ij}} 
\end{equation}
where the boundary is placed at $\rho=\epsilon$. This is related to the CFT stress tensor (which is the true renormalized stress tensor, and the one we are using in this section) through
\begin{eqnarray}
T_{ij} &=& \lim_{\epsilon \rightarrow 0}\left(\frac{1}{\epsilon^{d/2 -1}} T^{ren}_{ij}[\gamma] \right) = \lim_{\epsilon \rightarrow 0}\left(-\frac{2}{\sqrt{-g(x,\epsilon)}}\frac{\delta S^{ren}_D}{\delta g^{ij}} \right) \\ \nonumber &=& -\frac{2}{\sqrt{g_{0}}}\frac{\delta S^{ren}_D}{\delta g^{ij}_{0}}.
\end{eqnarray}
Here, $g_0$ is the leading term in the Fefferman-Graham expansion.}.
Remarkably, Compere-Marolf found that such boundary metric fluctuations are in fact normalizable with respect to the canonical (symplectic) structure defined by the {\em full} renormalized Dirichlet action $S^{ren}_D$. Furthermore they also showed that the symplectic structure is also conserved when the boundary condition $T_{ij}=0$ holds. They further showed that if we couple the full renormalized bulk Dirichlet action above to a boundary action that is a functional of the boundary metric (ie., the boundary is dynamical), so that the variation now becomes
\begin{eqnarray}
\delta S^{bndry}_D &\equiv & \delta \left(S_D + S_{bndry} \right) = \text{e.o.m} - \frac{1}{2}\int_{\partial \mathcal{M}} d^d x \sqrt{-g_{0}}T^{ij}\delta g_{0\;ij} \\ \nonumber &+& \int_{\partial \mathcal{M}} d^d x \frac{\delta S_{bndry}}{\delta g_{0\;ij}}\delta g_{0\;ij} \\ \nonumber &=& \text{e.o.m} -\frac{1}{2}\int_{\partial \mathcal{M}} d^d x \sqrt{g_{0}}\left( T^{ij} - \frac{2}{\sqrt{g_{0}}} \frac{\delta S_{bndry}}{\delta g_{0\;ij}}  \right)\delta g_{0\;ij}
\end{eqnarray}
then {\em again} the claims above hold, if instead of requiring $T^{ij}=0$ we now require 
\begin{equation}
T^{ij} - \frac{2}{\sqrt{g_{0}}} \frac{\delta S_{bndry}}{\delta g_{0\;ij}} = 0.\label{bounddyn}
\end{equation}

With that aside, let us turn to our Neumann case. We will merely discuss some connections between our work and and that of Compere-Marolf and leave it at that for now. We first note that the usual Dirichlet action plus a boundary term, after a Legendre transform of the kind we discussed, takes the form
\begin{eqnarray}
S^{bndry}_N &\equiv & S_D + S_{bndry} + \int_{\partial \mathcal{M}} d^d x \frac{\sqrt{g_{0}}}{2}\left( T^{ij} - \frac{2}{\sqrt{g_{0}}} \frac{\delta S_{bndry}}{\delta g_{0\;ij}}  \right)g_{0\;ij}
\end{eqnarray}
This has the variation
\begin{eqnarray}
\delta S^{bndry}_N &=& \text{e.o.m} + \frac{1}{2} \int_{\partial \mathcal{M}} d^d x  g_{0\;ij}\ \delta \left[ \sqrt{g_{0}} \left( T^{ij} - \frac{2}{\sqrt{g_{0}}} \frac{\delta S_{bndry}}{\delta g_{0\;ij}} \right)   \right]
\end{eqnarray}
Note that this is of the Neumann form, but now with boundary dynamics. It would be interesting to see if this leads to normalizable fluctuations, perhaps if one imposes the  condition (\ref{bounddyn}) that Compere and Marolf do. It is worth mentioning here that what \cite{Geoffrey} calls Neumann boundary condition is (as is often conventional in the gravity literature) the vanishing of  $T_{ij}$. This is the gravitational analogue of starting with the standard {\em Dirichlet} action in particle mechanics, letting the coordinate $q$ fluctuate at the boundary, but demanding that $\dot q=0$ at the boundary so that the boundary piece dies anyway, so that variational problem is well-defined. A genuinely Neumann condition is less constraining: it merely says that the normal derivative/canonical conjugate is {\em fixed}, not necessarily zero. This is what we do in this paper.

Of course to conclusively settle this question requires further work, but we suspect that when one takes into account the full dynamics of the system instead of a fixed AdS background, more boundary conditions than what are usually considered lead to consistent quantum theories. It seems likely that one can discuss the normalizability via the symplectic structure in a covariant phase space approach, and we will report on work in this direction elsewhere\footnote{Progress has been made in this direction after the first version of this paper appeared, it will be reported in \cite{CKCG}.}.


\section{Microcanonical in Gravity, Microcanonical in CFT, and Neumann}

The Neumann path integral that we have considered in this paper is related to the ``microcanonical" path integral that was considered by Brown-York \cite{brown}. Their approach amounts to holding {\em some} of the components of the quasi-local (boundary) stress tensor density fixed, whereas our approach is in some sense more covariant: we hold the entire boundary stress tensor density fixed. We saw that this has a natural interpretation as a Neumann problem, and results in a very simple Neumann action that leads to various nice features, some of which we investigated in \cite{KR, latest} as well as this paper. 

The path integral of \cite{brown} was called a ``microcanonical" functional integral. The motivation of \cite{brown} for this nomenclature was that in gravity, the total charges reduce to surface integrals over the boundary. In \cite{brown} this surface integral is not explicitly done, but we believe this surface integral actually needs to be done in order to get a true charge, and to make the path integral truly ``microcanonical" from the gravity perspective. 

We would like to emphasize however that even keeping the integrated charge (energy) fixed on the gravity side in the sense of \cite{brown} is not quite the same as holding the CFT energy fixed in AdS/CFT. This is because in \cite{brown} the boundary metric is allowed to fluctuate. In AdS/CFT however, in the microcanonical ensemble when we hold the CFT energy fixed, we also hold the metric fixed. If we have infinite resolution, there is no ensemble of states in the CFT satisfying both these conditions. 

In AdS/CFT, the natural microcanonical object to hold fixed from the CFT perspective is the total CFT energy, which should be compared to a charge (the boundary stress tensor density is a current from the CFT perspective). In the thermodynamic limit, the microcanonical density of states is a Laplace transform of the canonical partition function \cite{Chaichian}. The usual discussion of Hawking-Page transition in AdS/CFT is in the context of the canonical ensemble, but by doing this Laplace transform we can move to the microcanonical ensemble as well. The resulting discussion is guaranteed to match with the discussion of AdS thermodynamics in the microcanonical ensemble done in the Hawking-Page paper \cite{HawkPage}\footnote{This discussion is in the last section of their paper, and is not as well-known as their canonical discussion. The only thing relevant for our purposes here is that they change ensembles via the aforementioned Laplace transform.}, because the corresponding canonical discussions match. 

Our construction, as we have emphasized, is different from both \cite{brown} as well as the AdS/CFT discussion. Morally it is more similar to \cite{brown} because we also do not pin down the metric at the boundary. Our approach could be viewed as an alternate implementation of holography in AdS where the boundary metric is allowed to fluctuate. In a follow-up paper \cite{CKCG}, further evidence  will be provided (along the lines of the suspicions expressed in section 5) that these boundary conditions may be consistent boundary conditions for quantum gravity in AdS: we will find that in odd $d$ the fluctuations are normalizable, and that in even $d$, normalizability of the bulk fluctuations is guaranteed when the dynamics of the  boundary metric is controlled by conformal gravity. Another direction that is being explored is the possibility of doing a similar renormalized construction for flat space Neumann gravity along the lines of the Dirichlet case discussed by Mann and Marolf \cite{MannMarolf}. We have recently also constructed Robin boundary terms for gravity. Considering the fact that the Dirichlet boundary term \cite{York, GHY} has had numerous applications since its inception more than 40 years ago, perhaps it is not surprising that the Neumann term \cite{KR} also leads to natural applications and generalizations.

\section*{Acknowledgments}

We thank Pallab Basu, Justin David, Jarah Evslin, Bob McNees, Shahin Sheikh-Jabbari, Kostas Skenderis and Sergey Solodukhin for questions, comments and/or discussions. CK thanks the organizers and attendees at conferences at Bangalore, Phitsanulok and Yerevan, for discussions and hospitality.

\appendix

\section{Asymptotic solution}

The relation between the various $g_i$'s (with $i<d$) in Fefferman-Graham expansion is determined by solving Einstein's equation iteratively. This was worked out in detail in \cite{deHaro} and here we collect some useful results for completeness. The indices below are raised with the metric $g_{0}$.

The determinant of induced metric on $\rho=\epsilon$ boundary can be expanded as follows

\begin{equation}\label{detexpansion}
\sqrt{-g} = \sqrt{-g_0}\left( 1 + \frac{1}{2}\epsilon \Tr (g_{0}^{-1}g_2) + \frac{1}{8}\epsilon^2 \left( (\Tr (g_{0}^{-1}g_2))^2 - \Tr (g_{0}^{-1}g_2)^2 \right) + O(\epsilon^3) \right)
\end{equation}

The leading coefficients $g_n$ for $n\neq d$ are given by \footnote{Our convention for Ricci tensor and Ricci scalar differ from \cite{deHaro} by a minus sign}

\begin{eqnarray}
g_{2\; ij} &=& -\frac{1}{(d-2)}\left( R_{ij} - \frac{1}{2(d-1)}R g_{0\;ij} \right) \\ \nonumber
g_{4\; ij} &=& \frac{1}{(d-4)}\left( \frac{1}{8(d-1)}D_i D_j R - \frac{1}{4(d-2)}D^k D_k R_{ij} + \frac{1}{8(d-1)(d-2)}g_{0\; ij}D^k D_k R \right. \\ \nonumber  &-& \left. \frac{1}{2(d-2)}R^{kl}R_{ikjl} + \frac{(d-4)}{2(d-2)^2}R_{i}^{\; k}R_{kj} + \frac{1}{(d-1)(d-2)^2}R R_{ij} \right. \\ &+& \left. \frac{1}{4(d-2)^2}R^{kl}R_{kl}g_{0\;ij} - \frac{3d}{16(d-1)^2 (d-2)^2}R^2 g_{0\;ij}   \right)
\end{eqnarray}

For $n=d$, one can obtain the trace and divergence of $g_n$ as well as the coefficient of logarithmic term $h_d$ from Einstein's equation and we refer the reader to Appendix A of \cite{deHaro}. On-shell $g_2$ is determined in terms of the induced metric $\gamma$ as \cite{deHaro}

\begin{eqnarray}\label{g2traces}
\Tr\; g_2 &=& \frac{1}{2\epsilon (d-1)}\left( -R[\gamma] + \frac{1}{(d-2)}\left(R_{ij}[\gamma]R^{ij}[\gamma] - \frac{1}{2(d-1)}R^2[\gamma] \right) + O[R^3[\gamma]] \right) \nonumber \\ \\ \nonumber
\Tr\; g_{2}^{2} &=& \frac{1}{(d-2)^2 \epsilon^2}\left( R_{ij}[\gamma]R^{ij}[\gamma] + \frac{4-3d}{4(d-1)^2}R^{2}[\gamma] + O[R^3[\gamma]] \right)
\end{eqnarray}

\section{Legendre Transform Approach}

The Neumann action can be thought of as a boundary Legendre transform of the Dirichlet action. The Neumann and Dirichlet action are related by \cite{latest}

\begin{equation}\label{Neu_leg_action}
S^{ren}_N = S^{ren}_D - \int_{\partial \mathcal{M}}d^{D-1}x\;\pi^{ren}_{ab}\gamma^{ab}
\end{equation}
where $\pi^{ren}_{ab} = \frac{\delta S^{ren}_D}{\delta \gamma^{ab}}$. $\pi^{ren}_{ab}$ is further related to the renormalized boundary stress tensor as

\begin{equation}\label{pi-stress-tensor}
\pi^{ren}_{ab} = -\frac{\sqrt{-\gamma}}{2}T^{ren}_{ab}
\end{equation}
So, given the renormalized action and the boundary stress tensor for the Dirichlet case, we can use the above relations between the Dirichlet and Neumann action to obtain a renormalized action for the Neumann case. This serves as an independent check of the holographic renormalization of Neumann case and we will go through each case ($D=3,4,5$) separately here.

\subsection{AdS$_3$}
The renormalized Dirichlet action and stress-tensor for AdS$_3$ are given by \cite{balasubramanian, deHaro}

\begin{eqnarray}
S^{ren}_D &=& \frac{1}{2\kappa}\int_{\cal M} d^3x \sqrt{-g}(R-2\Lambda)+ \frac{1}{\kappa}\int_{\partial \mathcal{M}}d^2x\sqrt{-\gamma}\Theta \\ \nonumber &-& \frac{1}{\kappa}\int_{\partial \mathcal{M}}d^2x\sqrt{-\gamma}
\end{eqnarray}
and

\begin{equation}
T^{ren}_{ab} = \frac{1}{\kappa}\left( \Theta_{ab} - \Theta \gamma_{ab} + \gamma_{ab} \right)
\end{equation}
where we have set $l=1$. Using \eqref{Neu_leg_action} and \eqref{pi-stress-tensor} we immediately see that

\begin{equation}
S^{ren}_{N} = \frac{1}{2\kappa}\int_{\cal M} d^3x \sqrt{-g}(R-2\Lambda)+ \frac{1}{2\kappa}\int_{\partial \mathcal{M}}d^2x\sqrt{-\gamma}\Theta
\end{equation}
which matches with the renormalized Neumann action obtained by holographic renormalization.

\subsection{AdS$_4$}
In AdS$_4$, the renormalized Dirichlet action and stress tensor is \cite{balasubramanian, deHaro} (for $l=1$)

\begin{eqnarray}
S^{ren}_D &=& \frac{1}{2\kappa}\int_{\cal M} d^4x \sqrt{-g}(R-2\Lambda)+ \frac{1}{\kappa}\int_{\partial \mathcal{M}}d^3x\sqrt{-\gamma}\Theta \\ \nonumber &-& \frac{2}{\kappa}\int_{\partial \mathcal{M}}d^3x\sqrt{-\gamma}\left(1+ \frac{{}^{(3)}R}{4} \right)
\end{eqnarray}
and

\begin{equation}
T^{ren}_{ab} = \frac{1}{\kappa}\left( \Theta_{ab} - \Theta \gamma_{ab} + 2\gamma_{ab} - {}^{(3)}G_{ab} \right)
\end{equation}
Using \eqref{Neu_leg_action} and \eqref{pi-stress-tensor} we obtain

\begin{equation}
S^{ren}_N = \frac{1}{2\kappa}\int_{\cal M} d^4x \sqrt{-g}(R-2\Lambda) + \frac{1}{\kappa}\int_{\partial \mathcal{M}}d^3x\sqrt{-\gamma}\left(1- \frac{{}^{(3)}R}{4} \right)
\end{equation}
which is in agreement with the renormalized Neumann action obtained by holographic renormalization.

\subsection{AdS$_5$}
For the case of AdS$_5$ the renormalized action and stress-tensor are given by \cite{balasubramanian, deHaro} (for $l=1$)

\begin{eqnarray}
S^{ren}_D &=& \frac{1}{2\kappa}\int_{\cal M} d^5x \sqrt{-g}(R-2\Lambda)+ \frac{1}{\kappa}\int_{\partial \mathcal{M}}d^4x\sqrt{-\gamma}\Theta \\ \nonumber &-& \frac{3}{\kappa}\int_{\partial \mathcal{M}}d^4x\sqrt{-\gamma}\left(1+ \frac{{}^{(4)}R}{12} \right)
\end{eqnarray}
and 

\begin{equation}
T^{ren}_{ab} = \frac{1}{\kappa}\left( \Theta_{ab} - \Theta \gamma_{ab} + 3\gamma_{ab} - \frac{1}{2}{}^{(4)}G_{ab} \right)
\end{equation}
Using \eqref{Neu_leg_action} and \eqref{pi-stress-tensor} we once again obtain the renormalized Neumann action which matches with the one obtained by holographic renormalization

\begin{eqnarray}
S^{ren}_N &=& \frac{1}{2\kappa}\int_{\cal M} d^5x \sqrt{-g}(R-2\Lambda)- \frac{1}{2\kappa}\int_{\partial \mathcal{M}}d^4x\sqrt{-\gamma}\Theta \\ \nonumber &+& \frac{3}{\kappa}\int_{\partial \mathcal{M}}d^4x\sqrt{-\gamma}
\end{eqnarray}

\section{Gauss-Codazzi-Ricci relations}

Gauss-Codazzi relations helps us express the spacetime curvature tensors in terms of the intrinsic and extrinsic curvatures of the embedding hypersurface. They can be summarized as follows:

\begin{eqnarray}
R + 2R_{ab}u^a u^b &=& {}^{(2)}R - \hat{k}_{ab}\hat{k}^{ab} + \hat{k}^2 \\ \nonumber
\sigma_{ab}u_{c}R^{bc} &=& d_a \hat{k} - d_b \hat{k}^{b}_{a}  \\ \nonumber
\sigma_{ac}\sigma_{bd}R^{cd} &=& -\frac{1}{N}{\cal L}_{m}\hat{k}_{ab} - \frac{1}{N}d_a d_b N + {}^{(2)}R_{ab} + \hat{k}\hat{k}_{ab} - 2\hat{k}_{ac}\hat{k}^{c}_{b}
\end{eqnarray}
where ${\cal L}_m$ refers to the Lie derivative with respect to the vector $m^a = Nu^a$, $d_a$ is the covariant derivative w.r.t the metric $\sigma_{ab}$ and $\hat{k}_{ab}$ is the extrinsic curvature of $B$ embedded in ${\cal B}$. 

The last of these relations does not arise as commonly as the first two, we refer the reader to \cite{G-C-ref}. We need all three of them in our simplifications of the ADM version of the renormalized actions.

\end{document}